\documentclass[final]{siamltex}
\usepackage{xcolor}
\usepackage{amsmath,amssymb,amsfonts}
\usepackage{mathrsfs}
\usepackage{graphicx}
\usepackage{bm}
\usepackage{ucs}
\usepackage[utf8x]{inputenc}
\usepackage{wrapfig}
\usepackage{color}
\usepackage{float}
\usepackage{epstopdf}
\usepackage{algorithm,algorithmic}
\usepackage[breaklinks,colorlinks=true,linkcolor=blue,citecolor=red,
  backref=page]{hyperref}

\DeclareMathAlphabet{\itbf}{OML}{cmm}{b}{it}
\newtheorem{remark}{Remark}[section]
\renewcommand{\hat}{\widehat}
\renewcommand{\tilde}{\widetilde}

\newcommand{\RR}{\mathbb{R}}

\newcommand{\EE}{\mathbb{E}}

\newcommand{\bx}{{\itbf x}}
\newcommand{\by}{{\itbf y}}

\newcommand{\bk}{\boldsymbol{\kappa}}

\def\byp{y_\bot}

\def\bz{\itbf z}
\def\bzl{z_{\parallel}}
\def\bzp{z_\bot}

\def\byl{y_\parallel}

\def\bkp{\kappa_\bot}
\def\bkl{\kappa_\parallel}

\def\xp{x_\perp}

\def\om{\omega}
\def\la{\lambda}

\begin{document}
\title{High-Resolution Interferometric Synthetic Aperture Imaging in scattering media}
\author{ Liliana Borcea\footnotemark[1] \and Josselin
  Garnier\footnotemark[2] }
\renewcommand{\thefootnote}{\fnsymbol{footnote}}
\footnotetext[1]{Department of Mathematics, University of Michigan,
  Ann Arbor, MI 48109. {\tt borcea@umich.edu}} \footnotetext[2]{CMAP, CNRS, Ecole Polytechnique, Institut Polytechnique de Paris, 91128
  Palaiseau Cedex, France.  {\tt josselin.garnier@polytechnique.edu}}
\maketitle

\begin{abstract}
The goal of synthetic aperture imaging is to  estimate the reflectivity of a remote region of interest by processing  
data gathered with a moving sensor which emits periodically a signal   and records the backscattered wave.
We  introduce and analyze a high-resolution interferometric method  for synthetic aperture imaging  through an unknown scattering  medium which distorts the  wave.  The method builds on the coherent interferometric (CINT) approach which uses empirical cross-correlations of the measurements to mitigate the distortion, at the expense of a loss of resolution of the image. The new method shows that, while mitigating the wave distortion, it  is possible to obtain a  robust and sharp estimate of the modulus of the Fourier transform of the reflectivity function. A high-resolution image can then be obtained by a phase retrieval algorithm.
\end{abstract}

\section{Introduction}
\label{sec:intro}
\begin{figure}
\centerline{ \includegraphics[width=7.cm]{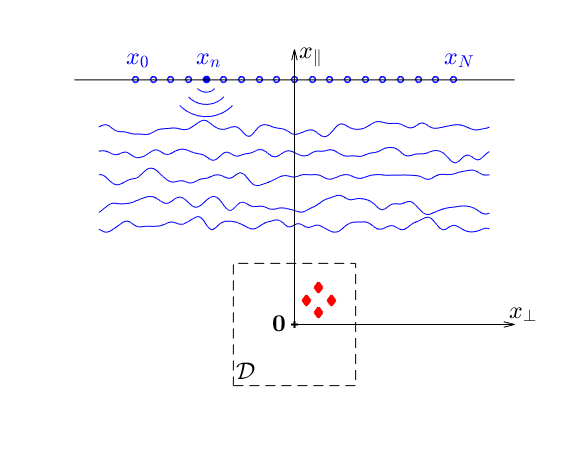} }
\vspace{-0.2in}\caption{Synthetic aperture imaging setup, where the sensor has the  successive positions  $(\bx_n)_{n=0}^N$ in a  linear aperture. The imaging region ${\cal D} $ which supports the reflectivity is centered at the origin ${\bf 0}$, the range coordinate is 
denoted by $x_{\parallel}$ and the cross-range by $\xp$. 
The medium between the aperture and the imaging region is randomly heterogeneous. }
\label{fig:dessinsar}
\end{figure}

In synthetic aperture imaging, a sensor  mounted on a moving platform (e.g., plane, satellite) emits  a signal $s(t-nT)$  at time instants $n T$, counted by  $n = 0, 1, \ldots, N$, and records the backscattered wave, the ``response"   $R_n(t)$.  The goal of imaging is to estimate the reflectivity $\rho(\bx)$ of a remote region of interest from  
\begin{equation}
\mbox{data} = \{R_n(t)  \mbox{ for } t\in (0,T), ~ n = 0, \ldots, N\}.
\label{eq:data}
\end{equation}
The signal $s(t)$ is  either a broad-band pulse defined by an envelope of small temporal support  of order $1/B \ll T$, where $B$ is the bandwidth, modulated at carrier frequency $\om_o \gg B$, or a chirp that can be compressed to a pulse with data processing \cite{curlander}. We assume the former. The variables $t$ and $nT$ are 
referred  to  as  the  ``fast  time"  and  the ``slow  time", respectively. The trajectory of the platform can be arbitrary, but for simplicity, and without loss of generality, we suppose that  it is straight and the motion is uniform, so that the signal emission is from the regularly spaced positions $\bx_n$, for  $n=0,\ldots,N$ (see Fig. \ref{fig:dessinsar}). The line segment connecting $\bx_0$ to $\bx_N$ is called the synthetic aperture and its length  $a = |\bx_N-\bx_0|$ is the aperture size. 

The data \eqref{eq:data} have only two degrees of freedom, so it is not possible to estimate  a reflectivity function $\rho(\bx)$ in three dimensions. Here we consider the problem in two dimensions, but the results extend to imaging in three dimensions on a surface with known topography. Again, for simplicity, we center the aperture above the remote region ${\cal D}$ of interest, and we introduce the system of coordinates $\bx = (x_\parallel,\xp)$ with origin at the center of ${\cal D}$, the  ``range coordinate" $x_{\parallel}$ measured  along the main direction of propagation, orthogonal to the aperture, and ``cross-range coordinate" $\xp$ measured along the aperture.

The classic synthetic aperture imaging method assumes that the medium between the sensor and the imaging region is non-scattering, with smooth and known wave speed. Without  going into technical details, the imaging function is given roughly by the sum over $n$ of the returns $R_n(t)$ evaluated (synchronized) at the roundtrip  travel time  between $\bx_n$ and the imaging point \cite{cheney,curlander}.  
When this point lies in the support of $\rho(\bx)$, denoted by $\mbox{supp}(\rho)$, the synchronized returns add constructively and the imaging function is large. Therefore,  the set $\mbox{supp}(\rho)$ can be estimated from the imaging function  displayed in ${\cal D}$ above some user defined threshold value. In the particular case of a homogeneous medium  with constant wave speed $c$,  and for an idealized reflectivity supported at two points, well known resolution formulas state that  these points can be distinguished  if they are separated by a distance of  order $c/B$ in range and $\la_o L/a$  in cross-range, where $\la_o = 2 \pi c/\om_o$ is the carrier wavelength and $L$ is the range offset from the aperture.

We are interested in imaging in heterogeneous media with microstructure, as sketched in  Fig. \ref{fig:dessinsar}, where the wave speed fluctuates 
about a  known reference profile, which we take equal to the constant $c$ for simplicity. The fluctuations have  small amplitude and occur on a length scale that is much smaller than $L$.  They are unknown and cannot be estimated as part of imaging from the band limited data \eqref{eq:data}. Thus, there is uncertainty in the wave propagation, which motivates modeling the wave speed as a random perturbation of the reference $c$. This model introduces  a stochastic framework where we can quantify the robustness of imaging methods with respect to the uncertainty of the microstructure.  Robust images cannot be obtained by empirical averaging over many realizations of the random wave speed, because the imaging experiment occurs in a single medium. However, with careful data processing, it is possible to get images that are practically insensitive to the particular realization i.e., are statistically stable.

The cumulative scattering effect of the microstructure  (the wave distortion) depends in a complicated way on the amplitude of the fluctuations of the wave speed as well as  the  relation between the length scale of the fluctuations, the carrier wavelength and the travel distance. When the distortion is strong, the classic synthetic aperture imaging method 
\cite{cheney,curlander} gives noisy images that are difficult to interpret and unreliable (statistically unstable). Different data processing
is needed to mitigate the wave distortion, based on the empirical cross-correlations of the measurements. 
The coherent interferometric (CINT) method  \cite{borcea11,borcea06} forms an image using such cross-correlations. It is known \cite{ishimaru,rytov,tatarski,vanRoss} that scattering causes statistical decorrelation of the time-harmonic components of the wave field over a frequency offset $\Omega_d$ called ``decoherence frequency" and a spatial offset $X_d$ called ``decoherence length". These scales depend on the statistics of the fluctuations of the wave speed, not the particular realization, and CINT takes them  into account  by calculating the empirical cross-correlations  in a time window of duration $1/\Omega$ and for sensor locations that are within a distance $X$ of each other. The image is then formed by superposing the empirical cross-correlations synchronized relative to the imaging point with travel time delays calculated in the reference medium. There is a trade-off between the resolution of the image and its robustness to the uncertainty of the microstructure, which is  quantified by the threshold parameters $X$ and $\Omega$ \cite{borcea11,borcea06}.
The smaller these  are, the less sensitive is the CINT image to the microstructure, as long as the aperture  and the bandwidth are large enough. However,  the range resolution  is of the order $c/\Omega $ and the cross-range resolution is of the order of $\la_o L/X $, so robustness comes at the cost of loss of resolution. The optimal choice is $X \approx X_d < a$ and $\Omega \approx \Omega_d < B$, and 
in practice this can be determined  by optimizing a measure of quality of the image \cite{borcea06}. 

CINT  has been used  for imaging  with arrays of sensors  \cite{borcea11,borcea06} (see also references therein)  and with synthetic apertures \cite{garniersolna08}. A modification of CINT introduced recently in \cite{borcea18} in the context of imaging  a constellation of point  sources with a passive array of receivers shows that it is possible  to localize the sources with resolution that is comparable to that in the homogeneous medium. More precisely,  nearby point sources  within a blurry peak of the CINT function can be resolved with resolution $c/B$ in range and $\la_o L/a$ in cross-range, up to an overall (rigid body) translation and rotation of the constellation. The algorithm in \cite{borcea18} involves a point search (it is targeted toward imaging a constellation of points) and has prohibitive computational cost   for many sources. In this paper we extend the ideas in \cite{borcea18} to synthetic aperture imaging of a general reflectivity $\rho(\bx)$, based on a new HCINT imaging function,  where the acronym stands for high-resolution CINT.  We show that this function is on one hand robust to the uncertainty of the fluctuations of the wave speed and on the other hand it allows a precise estimate of the modulus of the Fourier transform of $\rho(\bx)$. A high-resolution image can then be obtained from this estimate using phase retrieval \cite{fienup78,fienup82,fienup87,fienup90,review15}.

Our mathematical analysis of HCINT  is based on a geometrical optics model of wave propagation 
through random media. This simple model accounts for wavefront distortion and allows an explicit quantification 
of robustness i.e., calculation of the variance of the imaging function. The HCINT method is not model specific, and it can also be analyzed with more complex  wave propagation models like in \cite{GarnierSolna}.

The paper is organized as follows: We begin in section \ref{sect:imag} with the mathematical formulation 
of the problem and the expression of the three imaging functions: classic synthetic aperture imaging, CINT and  
HCINT. The model of wave propagation in the random medium is described in section \ref{sect:model}
and it is used to analyze the imaging functions in  section \ref{sect:analImage}. The estimation of the 
modulus of the Fourier coefficients of the reflectivity function and the subsequent 
imaging based on phase retrieval is in section \ref{sect:phaseRet}. We present numerical results in section  \ref{sect:numerics} and  end with a summary in section \ref{sect:sum}.

\section{Formulation of the problem and the imaging functions}
\label{sect:imag}
The wave $u_n(t,\bx)$ emitted  from the location $\bx_n = (L,x_{n\perp})$ in the aperture satisfies
\begin{align}
\frac{1}{c^2}\Big[ 1 + \sigma \mu\Big(\frac{\bx}{\ell_c}\Big) + \rho(\bx) \Big] \partial_t^2 u_n(t,\bx) - \Delta u_n(t,\bx)&= s(t-nT) \delta(\bx-\bx_n), 
\label{eq:waveeq} 
\end{align}
for $n = 0, \ldots, N$, time $t \in \RR$ and position $\bx \in \RR^2$, with  the initial condition
\begin{align}
u_n(t,\bx) &\equiv 0, \quad t \in (-\infty, nT) \setminus {\rm supp} \big( s(t-nT)\big).\label{eq:quiet}
\end{align}
Here $c$ is the constant reference wave speed, $\mu$ models the random fluctuations and $\rho$ is the unknown reflectivity,  assumed compactly supported away from the aperture. The random process $\mu$ 
is statistically homogeneous, with 
mean zero and  integrable autocovariance
\begin{equation}
{\cal R}(\bx-\bx') = \EE \big[ \mu(\bx)\mu(\bx') \big] ,
\end{equation}
normalized so that
\begin{equation}
{\cal R}({\bf 0}) = 1  \mbox{ and }
\int_{\RR^2}  d \bx \, {\cal R}(\bx)   = 1 .
\end{equation}
We assume henceforth, for convenience and without loss of generality, the Gaussian autocovariance
\begin{equation}
{\cal R}(\bx) = \exp(-\pi |\bx|^2).
\label{eq:GaussCov}
\end{equation}
The dimensionless parameter $\sigma$ in  \eqref{eq:waveeq} is the standard deviation of the random fluctuations and 
the length scale $\ell_c$ is the correlation length. 

The inverse problem is to estimate the reflectivity $\rho$ from the data \eqref{eq:data}. We study its solution 
using the three imaging functions given in  sections \ref{sect:imSAR}--\ref{sect:HCINT}. Their expression is 
based on two standard approximations:  (1) the  ``start-stop" approximation \cite{cheney,curlander} which assumes that 
the sensor movement during the  roundtrip travel time to the imaging region ${\cal D}$ is negligible; (2) the single scattering (Born) approximation which assumes that the reflectivity $\rho$ is not too strong. 
The data model is 
\begin{align}
R_n(t) &=  u_n(t,\bx_n) + W_n(t) \nonumber \\
&=
\frac{1}{2\pi } \int_\RR  d \om \,   e^{-i \omega (t-nT)} \hat{s}(\omega) k^2(\om)
\int_{\RR^2} d \by \, \rho(\by) \hat{G}^2_\mu(\omega,\by,\bx_n)  +W_n(t),
\label{eq:datamodel}
\end{align}
where $k(\om) = \om/c$ is the wavenumber and we denote with ``hat" the Fourier transform with respect to time, defined 
with the convention 
$$
\hat{s}(\omega) = \int_\RR dt \, s(t) e^{i \omega t} ,\quad
 s(t)  =  \frac{1}{2\pi} \int_\RR d \om \, e^{- i \omega t} \hat s(\om).
$$
The propagation through the random medium is modeled in \eqref{eq:datamodel} with the 
Green's function $\hat G_\mu$ of the Helmholtz equation with wave speed 
$c \big[1 + \sigma \mu(\bx/\ell_c)\big]^{-1/2}$ and $W_n$ denotes additive noise. 
For convenience, we let $W_n(t)$ be  Gaussian, white in time $t$ and uncorrelated in $n$, 
with mean zero and covariance given in the frequency domain by 
\begin{equation}
\EE \big[ \overline{\hat{W}_n(\omega)} \hat{W}_{n'}(\omega') \big] = \sigma_{\rm W}^2 \delta(\omega - \omega')
%\delta(x_{n\perp}-x_{n'\perp}) 
\delta_{nn'} ,
\label{eq:covarW0}
\end{equation}
where $\delta_{nn'}$ stands for the Kronecker symbol.
The bar is used throughout to denote complex conjugate.

\subsection{Synthetic aperture radar (SAR) imaging}
\label{sect:imSAR}
The fluctuations of the wave speed are neglected in the standard synthetic aperture imaging method, meaning that
the wave propagation is modeled by the Green's function in the reference medium
\begin{align}
\hat{G}(\omega,\bx,\by) = \frac{i}{4} H_0^{(1)} \big( k( \omega)|\bx-\by|\big) 
\approx \frac{\exp\big( i k(\om) |\bx-\by| +i \frac{\pi}{4} \big)}{2^{3/2}  \sqrt{ \pi k(\omega) |\bx-\by|}},\label{eq:greenhomo}\end{align}
where $H_0^{(1)}$ is the Hankel function of the first kind and of order $0$ and the approximation is 
for a large distance 
$|\bx-\by|$ with respect to the wavelength $\la = 2 \pi /k(\om)$. 

The  imaging function 
\begin{align}
\nonumber
{\cal I}_{\rm{SAR}}(\by^S) &= \Big| \sum_{n=0}^N \int_\RR dt \, R_n(t)  \overline{F_n(t-nT,\by^S) }\Big|^2
\\
&=\Big| \frac{1}{2\pi}\sum_{n=0}^N \int_\RR d \om \, \hat{R}_n(\omega) \overline{\hat F_n(\omega,\by^S)} e^{ -i \omega nT} \Big|^2,
\label{eq:SAR}
\end{align}
is the superposition of the data \eqref{eq:data} convolved (matched filtered) with 
\begin{equation}
F_n(t,\by^S) = \frac{1}{2\pi} \int_{\RR} d \om \, e^{-i \om t} \hat F_n(\omega,\by^S), 
\quad \hat F_n(\omega,\by^S) =   \hat{s}(\omega) \hat{G}^2(\omega,\by^S,\bx_n).
\label{eq:filterH}
\end{equation}
Note that $F_n$ is the data model for a point reflector at the search (imaging) point $\by^S$ in the reference medium. 
The matched filtering is called henceforth ``backpropagation" to the imaging point $\by^S$. At long range
$|\bx_n-\by^S | \sim L \gg a$, where ``$\sim$" means of the order of, and neglecting constant amplitude factors,  the backpropagation amounts to evaluating the response $R_n(t)$ at the  roundtrip travel time $2 |\bx_n-\by^S|/c$.
The index SAR in \eqref{eq:SAR} is the acronym for synthetic aperture radar,  the most common application of the
synthetic aperture imaging modality.

\subsection{The CINT imaging function}
\label{sect:CINT}
The CINT imaging function is obtained by backpropagating selected empirical cross-correlations of the measured 
responses, for nearby pairs of sensor locations $\bx_n,\bx_{n'}$ and frequencies $\omega,\omega'$,
\begin{align}
\nonumber
{\cal I}_{\rm CINT}(\by^S) = &
\frac{1}{(2\pi)^2} \sum_{n,n'=0}^N \int_{\RR} d \om \int_{\RR} d \om' \, 
\overline{\hat{R}_n(\omega)} {\hat{R}_{n'}}(\omega')
{\hat F_n}(\omega,\by^S)\overline{\hat F_{n'}(\omega',\by^S)}\\
& \times
e^{i \omega nT-i\omega' n'T} \exp\Big( -\frac{|\bx_n-\bx_{n'}|^2}{2X^2}- \frac{(\omega-\omega')^2}{2\Omega^2} \Big).
\label{eq:CINT} 
\end{align}
We use Gaussian sensor offset and frequency windows for convenience in the calculations, with standard deviations $X$ and $\Omega$ accounting for the decorrelation of the wave components due to scattering,  as explained in the analysis in section \ref{sect:analImage}.

\subsection{The HCINT imaging function}
\label{sect:HCINT}
Instead of \eqref{eq:CINT}, consider the CINT-like imaging function defined for two nearby imaging points $\by^S$ and ${\by^S}'$, as proposed in a passive array imaging context in \cite{borcea18},
\begin{align}
\nonumber
{\cal I}(\by^S,{\by^S}') = &
\frac{1}{(2\pi)^2} \sum_{n,n'=0}^N \int_{\RR} d \om  \int_{\RR}
d \om' \, \overline{\hat{R}_n(\omega)} {\hat{R}_{n'}}(\omega')
{\hat F_n}(\omega,\by^S)\overline{\hat F_{n'}(\omega',{\by^S}')}\\
& \times
e^{i \omega nT-i\omega' n'T} \exp\Big( -\frac{|\bx_n-\bx_{n'}|^2}{2X^2}- \frac{(\omega-\omega')^2}{2\Omega^2} \Big).
\label{def:cintg}
\end{align}
We call it the ``two-point CINT" imaging function and note that it is a generalization of  the CINT imaging function (\ref{eq:CINT}) since we have
\begin{equation}
\label{eq:HCINT_CINT}
{\cal I}_{\rm CINT}(\by^S) = {\cal I}(\by^S,{\by^S}). 
\end{equation}
The HCINT function is defined by the integral of \eqref{def:cintg} over the center locations
\begin{equation}
{\cal I}_{\rm HCINT}(\tilde \by^S) = \int_{\RR^2} d {\by^S} \, {\cal I}\Big({\by^S}+\frac{\tilde \by^S}{2},{\by^S}-\frac{\tilde \by^S}{2}\Big).   
\label{def:HCINT}
\end{equation}
We will also use  its Fourier transform
\begin{align}
\widehat{\cal I}_{\rm HCINT}(\bk) &= \int_{\RR^2} d\tilde \by^S \, {\cal I}_{\rm HCINT}(\tilde \by^S)  e^{-i \bk\cdot \tilde \by^S } \nonumber \\ &=
\int_{\RR^2} d\by^S \int_{\RR^2} d{\tilde \by^S}   \,  {\cal I}\Big ({\by^S}+\frac{\tilde \by^S}{2},{\by^S}-\frac{\tilde \by^S}{2}\Big)   e^{-i \bk \cdot \tilde \by^S }.
\label{eq:HatHCINT}
\end{align}

We will show in sections \ref{sect:analImage}--\ref{sect:phaseRet} that  ${\cal I}_{\rm CINT}$  gives a statistically stable but low resolution image of the reflectivity $\rho$, whereas the  HCINT imaging function \eqref{eq:HatHCINT} gives an 
estimate of the modulus of the Fourier transform of  $\rho$. This estimate can then be used in a
phase retrieval algorithm \cite{fienup78,fienup82,fienup87,fienup90,review15} to get a high-resolution image of the reflectivity. 

In practice, HCINT may be used to improve the resolution by zooming a region of interest in the support of the CINT image. The integrals over ${\by^S}$ and $\tilde \by^S$ in \eqref{eq:HatHCINT} can be evaluated with numerical quadrature formulas in such a small zoom region.

% ----------------------------------------------
\section{Random travel time model}
\label{sect:model}
In this section we briefly review the geometrical optics model of wave propagation through the random medium with wave speed $c [ 1 + \sigma \mu(\bx/\ell_c)]^{-1/2}$. Its derivation is given in \cite[Section 12.1]{noisebook}
under the high-frequency scaling assumption $\la_o \ll \ell_c < L$ and the weak fluctuations assumption 
$\sigma^2 \ll (\ell_c/L)^3$, so that scattering does not effect the amplitude of the wave  and the rays remain straight. It is only the  travel time calculated  along the straight ray that is randomized. We are interested in a long range $L \gg \ell_c$, 
where the random travel time fluctuations  have Gaussian statistics (even if $\mu$ is not Gaussian), by the central limit theorem. To showcase the effect of the random medium, we assume that these fluctuations are large (wave front is strongly distorted), which amounts to having
\begin{equation}
\sigma^2 \frac{L^3}{\ell_c^3}  \ll
 \frac{\lambda_o^2}{\sigma^2 {\ell_c} L} \ll 1.
 \label{eq:condrtt}
\end{equation}

The Green's function  is
\begin{align}
\hat{G}_\mu(\omega,\bx,\by) &\approx \hat{G}(\omega,\bx,\by)  \exp \big( i \om {\cal T}_\mu(\bx,\by)\big), 
\label{eq:green1}
\end{align}
where $\hat G$ is given in \eqref{eq:greenhomo} and 
\begin{equation}
 {\cal T}_\mu({\bx},{\by}) = \frac{\sigma |{\bx}-{\by}|}{2c} \int_0^1  dh \, \mu \Big(
\frac{{\by} + h({\bx}-{\by})}{\ell_c} \Big)
\end{equation}
models the random fluctuations of the travel time, given by the line integral of the random process $\mu$ 
along the straight ray connecting $\bx$ and $\by$. For points  $\by,\by'$  in the neighborhood of the origin,
satisfying $|\by-\by'|<\ell_c$, and for  $\bx_n,\bx_{n'}$ in the aperture,
the process ${\cal T}_\mu $ has Gaussian statistics with mean zero 
and covariance function
\begin{equation}
\EE\big[ {\cal T}_\mu (\bx_n,\by){\cal T}_\mu(\bx_{n'},\by') \big] = \tau^2
{\cal C}\Big(\frac{|\bx_n-\bx_{n'}|}{\ell_c} \Big), \qquad {\cal C}(r) = \frac{1}{r} \int_0^r  d h \,  e^{-\pi h^2}.
\end{equation}
Here we used the assumption \eqref{eq:GaussCov} and introduced the time scale 
\begin{equation}
\tau = \frac{ \sigma \sqrt{\ell_c L}}{2 c},
\label{eq:defTau}
\end{equation}
which quantifies the standard deviation of the random fluctuations of the travel time.  Note that 
\begin{equation}
\om \tau \sim \om_o \tau \gg 1,
\label{eq:largeTau}
\end{equation}
by the assumption \eqref{eq:condrtt}, so the phase of the Green's function \eqref{eq:green1} has very large fluctuations.

For arbitrary four points $(\bx_{n_j})_{j=1,\ldots,4}$ in the aperture, indexed by $0 \le n_j \le N$, 
four points $(\by_j)_{j=1,\ldots,4}$ in the search (imaging) region ${\cal D}$ with diameter smaller than $\ell_c$, 
and for frequencies $(\omega_j)_{j=1,\ldots,4}$, we have by the Gaussian property of ${\cal T}_\mu$ 
that 
\begin{align}
\EE \big[ \exp\big( 2i \omega_1 {\cal T}_\mu (\bx_{n_1},\by_1) \big)  \big]
=&
 \exp \big( - 2 \omega^2  \tau^2  \big) , \label{eq:mean}\\
\nonumber
\EE \big[ \exp\big(  2i \omega_1 {\cal T}_\mu(\bx_{n_1},\by_1) -2 i \omega_2  {\cal T}_\mu(\bx_{n_2},\by_2)\big) \big]
=& \exp \Big\{ -2 (\omega_1-\omega_2 )^2 \tau^2  \\
&  \hspace{-0.8in}- 4\omega_1 \omega_2  \tau^2 \Big[ 1-
  {\cal C}\Big(\frac{|\bx_{n_1}-\bx_{n_2}|}{\ell_c}\Big) \Big] \Big\}, \label{eq:twomom}
\end{align}
and
\begin{align}
\nonumber
&\EE \big[ \exp\big( 2i \omega_1 {\cal T}_\mu(\bx_{n_1},\by_1) - 2i \omega_2 {\cal T}_\mu(\bx_{n_2},\by_2)
- 2 i \omega_3 {\cal T}_\mu(\bx_{n_3},\by_3) +2 i \omega_4 {\cal T}_\mu(\bx_{n_4},\by_4) \big) \big]\\
\nonumber
&
= \exp\bigg\{ -2 \tau^2 \Big[ \sum_{j=1}^4 \omega_j^2 +2
\omega_1\omega_4 {\cal C}\Big(\frac{|\bx_{n_1}-\bx_{n_4}|}{\ell_c}\Big) + 2
\omega_2\omega_3 {\cal C}\Big(\frac{|\bx_{n_2}-\bx_{n_3}|}{\ell_c}\Big)   \\
\nonumber
&\quad  -2 \omega_1\omega_2 {\cal C}\Big(\frac{|\bx_{n_1}-\bx_{n_2}|}{\ell_c}\Big)
-2 \omega_1\omega_3 {\cal C}\Big(\frac{|\bx_{n_1}-\bx_{n_3}|}{\ell_c}\Big) -2
\omega_2\omega_4 {\cal C}\Big(\frac{|\bx_{n_2}-\bx_{n_4}|}{\ell_c}\Big) \\
&\quad -2
\omega_3\omega_4 {\cal C}\Big(\frac{|\bx_{n_3}-\bx_{n_4}|}{\ell_c}\Big) \Big] \bigg\}.
\label{eq:momfour1}
\end{align}

We conclude from equations \eqref{eq:green1},  \eqref{eq:largeTau} and \eqref{eq:mean}  that 
\begin{equation}
\EE \big[ \hat G_\mu(\om,\by_j,\bx_{n_j} \big] \approx 0, \quad j = 1, \ldots, 4.
\end{equation}
Physically, this means that the wave front is strongly distorted (randomized)
due to scattering, so that averaging the Green's function over realizations of the wave speed gives a negligible result. We also get from \eqref{eq:twomom} and the assumption on the bandwidth   \[|\omega_{1,2} -\omega_o| \sim B \ll \om_o,\]  that  the second moments \eqref{eq:momfour1} are negligible unless $\bx_{n_1}$ and $\bx_{n_2}$ are nearby. For such points we can expand the covariance in the phase of \eqref{eq:twomom} around the origin and obtain the simpler formula
\begin{align}
\EE\big[ \hat G_\mu(\om_1,\by_1,\bx_{n_1})\overline{\hat G_\mu(\om_2,\by_2,\bx_{n_2})}\big]
&\approx  \hat G(\om_1,\by_1,\bx_{n_1})\overline{\hat G(\om_2,\by_2,\bx_{n_2}) }
\nonumber \\
&\times \exp\left[ - \frac{|\bx_{n_1}-\bx_{n_2}|^2}{2X_d^2} - \frac{(\omega_1-\omega_2)^2}{2\Omega_d^2}\right].
\label{eq:secondmomentgreen}
\end{align}
The scales of decay in the sensor  and frequency offsets
\begin{equation}
X_d = \frac{\sqrt{3} \la_o \sqrt{\ell_c}}{(2 \pi)^{3/2} \sigma \sqrt{L}}  ,
 \quad \quad 
\Omega_d = \frac{1}{2\tau} =\frac{ c}{\sigma \sqrt{\ell_c L}}.
\label{eq:defXc}
\end{equation}
quantify the statistical decorrelation of the wave in the random medium and  are called the decoherence length and  decoherence  frequency.

% ---------------------------------
\section{Analysis of the imaging functions}
\label{sect:analImage}
We now use  the random travel time model of wave propagation to analyze the three imaging functions given in sections \ref{sect:imSAR}--\ref{sect:HCINT}. We compare the results to those in the reference homogeneous medium and account 
for the effects of the additive noise, as well. 

In the analysis we choose a probing pulse with Gaussian envelope
\begin{equation}
\label{eq:pulse}
s(t) =  e^{-i \omega_o t} \frac{B}{\sqrt{2\pi}}  \exp \Big( -\frac{B^2 t^2}{2}\Big) ,
\end{equation}
and with Fourier transform 
\begin{equation}
\hat s(\om) = \int_{\RR} dt \, s(t) e^{i \om t} = \exp \Big[ -\frac{(\om-\om_o)^2}{2B^2}\Big].
\label{eq:hatpulse}
\end{equation}
The successive positions of the sensor are close to each other
\[
|\bx_{n+1}-\bx_n| = \frac{a}{N} \ll a, \qquad n = 0, \ldots, N-1,
\]
so we approximate the sums over the index $n $ of $\bx_n$  by integrals over the aperture
and take the Gaussian appodization $\exp(-\xp^2/a^2)$
\begin{equation}
\label{eq:contAp}
\sum_{n = 0}^N \leadsto \int_{\RR} d \xp e^{-{\xp^2}/{a^2}}.
\end{equation}
The aperture size $a$ is assumed smaller than the range $L$, so we can use the paraxial approximation
\begin{equation}
|\bx - \by^S| \approx L-\byl^S + \frac{(\xp-\byp^S)^2}{2 L},
\label{eq:paraxAp}
\end{equation}
for all points $\bx = (L,\xp)$ in the aperture and points $\by^S = (\byl^S,\byp^S)$ in the imaging region ${\cal D}$.
These choices lead to explicit expressions of the imaging functions, but do not play an essential role in the conclusions.

\subsection{Analysis of the SAR imaging function}
\label{sect:analSAR}
The index $n$ is no longer needed in the continuum aperture approximation \eqref{eq:contAp},
so we change slightly the notation 
\[
\hat R_n(\om)e^{-i \om n T} \leadsto \hat R(\om,\xp), \quad \hat F_n(\om,\by^S) \leadsto \hat F(\om,\xp,\by^S), 
\quad \hat W_n(\om) \leadsto \hat W(\om,\xp).
\]
From now on, $W(t,\xp)$ is a Gaussian process, white in time $t$ and in space $\xp$, 
with mean zero and covariance given in the frequency domain by 
\begin{equation}
\EE \big[ \overline{\hat{W}(\omega,\xp)} \hat{W} (\omega',\xp') \big] = \sigma_{\rm W}^2 \delta(\omega - \omega')
\delta(\xp-\xp')  .
\label{eq:covarW}
\end{equation}
The expression \eqref{eq:SAR} of the imaging function becomes
\begin{align}
\label{eq:SAR1}
{\cal I}_{\rm SAR}(\by^S) 
&=\Big| \frac{1}{2\pi}\int_\RR d \om  \int_\RR d \xp \hat{R}(\omega,\xp) \overline{\hat{F}(\omega,\xp,\by^S)} e^{- \xp^2/a^2}\Big|^2,
\end{align}
with 
\begin{align}
\hat R(\om,\xp) =  \hat s(\om) k^2(\om)
\int_{\RR^2} d \by \, \rho(\by) \hat{G}^2_\mu(\omega,\by,(L,\xp))  +\hat W(\om,\xp),
\label{eq:hatR}
\end{align}
and 
\begin{equation}
\hat F(\om,\xp,\by^S) = \hat s(\om)\hat G^2(\om,\by^S,(L,\xp)), \qquad \hat s(\om) = \exp\Big[ -\frac{(\om-\om_o)^2}{2B^2}\Big] .
\label{eq:hatD}
\end{equation}

\subsubsection{Homogeneous medium}
In the absence of the fluctuations of the wave speed and noise,  
the expression  of the imaging function \eqref{eq:SAR1}  would be 
\begin{align}
{\cal I}_{\rm SAR}(\by^S) 
=  {\rm C} \Big|  \int_{\RR^2} d \by \, \rho(\by)  {\cal K}_{a,B}(\by^S-\by)  \Big|^2,  \quad {\rm C} = \frac{1}{2^{14} \pi^6 L^4},
\label{eq:SARhom}
\end{align}
where   $k_o = k(\om_o) = 2 \pi/\la_o$ and 
\begin{equation}
{\cal K}_{a,B}(\by) =  \pi a B  \exp\left[ - \frac{\byp^2}{[L/(k_o a)]^2}
-\frac{ \byl^2}{(c/B)^2} -2 i k_o \byl\right].
\label{eq:SARKern}
\end{equation}
This  is obtained  with straightforward calculations from equations \eqref{eq:greenhomo}, 
(\ref{eq:SAR1}--\ref{eq:hatD}) and the paraxial approximation \eqref{eq:paraxAp}.

In the particular case of a point reflector at  location $\by_\star$, the imaging function is proportional to the square modulus of the ``point-spread function" ${\cal K}_{a,B}$,
\begin{equation}
\label{eq:point scatterer}
 {\cal I}_{\rm SAR}(\by^S)  = {\rm C} \, \rho_\star^2 \, 
|{\cal K}_{ {a}, {B}}(\by^S -\by_\star)|^2, \quad \mbox{for}~~\rho(\by)=\rho_\star \delta(\by-\by_\star).
\end{equation}
This and the expression \eqref{eq:SARKern} show that  the resolution in the cross-range 
direction is  $\sim \lambda_o L/a$ and in the range direction is 
$\sim c/B$, as reported in the literature \cite{cheney}. 

\subsubsection{Random medium}
To explain the behavior of the SAR imaging function in the random medium, 
we describe here  its expectation and covariance in the noiseless case (i.e., when there is no additive noise). 
The  effect of the additive noise is analyzed in the next section.

We obtain from equations \eqref{eq:greenhomo}, 
(\ref{eq:paraxAp}--\ref{eq:hatD}) and the  moment formula \eqref{eq:secondmomentgreen} that 
\begin{align}
\nonumber
\EE\big[ {\cal I}_{\rm SAR}(\by^S) \big]
=&
{\rm C} \int_{\RR^2} d \by \int_{\RR^2} d \by' \, \rho(\by) \rho(\by'){\cal K}_{\tilde{a},\tilde{B}}(\by^S -\by)
{\overline{{\cal K}_{\tilde{a},\tilde{B}}(\by^S -\by')} }   \\
&\times
\exp \left[ - \frac{B^2 (\byl'-\byl)^2}{2 \Omega_d^2 (c/\tilde B)^2} 
- \frac{ a^2 (\byp'-\byp)^2}{2 X_d^2 [L/(k_o \tilde a)]^2} \right],
\label{eq:meanSAR}
\end{align}
with the same constant ${\rm C}$ as in \eqref{eq:SARhom} and with $\tilde a$ and $\tilde B$ defined by
\begin{equation}
\frac{1}{\tilde{a}^2} = \frac{1}{a^2}+\frac{1}{X_d^2},\quad \quad \frac{1}{\tilde{B}^2} = \frac{1}{B^2}+\frac{1}{\Omega_d^2} .
\label{eq:tildeaB}
\end{equation}
If  the random medium is strong enough so that the decoherence parameters \eqref{eq:defXc} satisfy $X_d < a$ and/or $\Omega_d<B$, we obtain from \eqref{eq:meanSAR} that the amplitude of expectation of the imaging function is reduced
and there is loss of resolution. This is evident in the case of a single  point scatterer, where 
\begin{equation}
\EE\big[ {\cal I}_{\rm SAR}(\by^S) \big] = {\rm C}\rho_\star^2 
{ |{\cal K}_{\tilde{a},\tilde{B}}(\by^S -\by_\star)|^2 }, \quad \mbox{for} ~~ \rho(\by)=\rho_\star \delta(\by-\by_\star).
\label{eq:mean point}
\end{equation}
Comparing with \eqref{eq:point scatterer} and using the definition \eqref{eq:SARKern}, we note that the peak amplitude of  \eqref{eq:mean point} is smaller by the factor 
$(\tilde{a}/a)^2(\tilde{B}/B)^2$ and the 
resolution is reduced to ${\lambda_o L }/{\tilde{a}}$ and 
 $c/\tilde{B}$  in the cross-range and range direction. These resolution limits are the scales of decay of the kernel  ${\cal K}_{\tilde{a},\tilde{B}}$ defined in  \eqref{eq:SARKern}, with $a$ and $B$ replaced by $\tilde{a}$ and $\tilde{B}$.

The calculation of the covariance of ${\cal I}_{\rm SAR}$ involves the fourth order moments \eqref{eq:momfour1},
for distinct points and  frequencies satisfying
\begin{eqnarray*}
&&\omega_1 = \omega + \frac{\tilde{\omega}}{2}, \quad \omega_2= \omega -
\frac{\tilde{\omega}}{2}, \quad \quad \omega_3 = \omega' + \frac{\tilde{\omega}'}{2},
\quad \omega_4= \omega' - \frac{\tilde{\omega}'}{2},\\
 &&\bx_1 = \bx + \frac{\tilde{\bx}}{2}, \quad \bx_2= \bx -
\frac{\tilde{\bx}}{2}, \quad \quad \bx_3 = \bx' + \frac{\tilde{\bx}'}{2}, \quad
\bx_4= \bx' - \frac{\tilde{\bx}'}{2}.
\end{eqnarray*}
Here we introduced the center frequencies $\omega,\omega' \sim \omega_o$ and the frequency offsets which satisfy $|\tilde{\omega}|, |\tilde{\omega}'|\lesssim 2 B$. Similarly, $\bx,
\bx' $ are center points in the aperture i.e., with range coordinate $L$ and cross-range coordinates in the interval $(-a/2,a/2)$. 
The spatial offsets  are $\tilde{\bx} = (0,\tilde x_\perp)$ and $ \tilde{\bx}' = (0,\tilde x_\perp')$, with  $|\tilde x_\perp|, 
|\tilde x_\perp'| \lesssim a$. 

If the diameter of the search region ${\cal D}$ is smaller than $\ell_c$ and if $X_d<a$ and $\Omega_d<B$,
then we find the covariance
\begin{align*}
{\rm Cov}\big( {\cal I}_{\rm SAR}(\by^S) , {\cal I}_{\rm SAR}({\by^S}') \big) &={\rm C}^2\Bigg|  \int_{\RR^2} d \by  \int_{\RR^2} d \by' \, \rho ( \by  ) \rho ( \by' ) {\cal K}_{\tilde{a},\tilde{B}}(\by^S -\by)
{ \overline{{\cal K}_{\tilde{a},\tilde{B}}({\by^S}' -\by')} } \\
& \hspace{-0.5in}\times
\exp\left[ - \frac{B^2 (\byl'-\byl-\byl^S+{\byl^S}')^2}{2 \Omega_d^2 (c/\tilde B)^2} 
- \frac{a^2 (\byp'-\byp-\byp^S+{\byp^S}')^2}{2X_d^2 [L/(k_o \tilde a)]^2} \right]\Bigg|^2.
\end{align*}
This indicates that the image displays bright and dark spots, so-called ``speckle" of size 
of the order of the correlation radius $\la_o L/\tilde{a}$ in the cross-range direction and $c/\tilde{B}$ in the range direction,
the scales of decay of the kernel $ {\cal K}_{\tilde{a},\tilde{B}}$.
Moreover, if we let $\by^S = {\by^S}'$ in this expression, we obtain that the variance is equal to the square mean:
\begin{equation}
{\rm Var}\big( {\cal I}_{\rm SAR}(\by^S) \big) = \EE\big[ {\cal I}_{\rm SAR}(\by^S) \big]^2.
\label{eq:SAR_VAR}
\end{equation}
Therefore, the coefficient of variation (i.e., the ratio of the standard deviation over the mean) of the SAR image near its peak values is large,
\begin{equation}
\mathcal{V}_{\rm SAR}(\by^S) = \frac{\sqrt{{\rm Var}\big( {\cal I}_{\rm SAR}(\by^S) \big)}}{\EE\big[ {\cal I}_{\rm SAR}(\by^S) \big] }= 1,
\label{eq:variationSAR}
\end{equation}
i.e., the SAR image has strong random fluctuations,  it is not statistically stable. 

\subsubsection{Additive noise}
The presence of the Gaussian additive noise  introduces an additional speckle pattern in the image, modeled by 
\begin{align}
\label{eq:SARW}
{\cal I}_{{\rm SAR},{\rm W}}(\by^S) 
&=\Big| \frac{1}{2\pi}\int_\RR d \om  \int_\RR d \xp \hat{W}(\omega,\xp) \overline{\hat{F}(\omega,\xp,\by^S)} e^{- \xp^2/a^2}\Big|^2.
\end{align}
This is independent of the fluctuations of the wave speed in the random medium.  
We now describe the mean and correlation radius of \eqref{eq:SARW}, where the latter gives  the typical noise induced speckle size. 

We obtain after straightforward calculations, using definitions \eqref{eq:covarW}, \eqref{eq:greenhomo}, \eqref{eq:filterH} 
and the paraxial approximation \eqref{eq:paraxAp} that the mean of the speckle pattern is uniform
\begin{equation}
\label{def:Cn}
\EE\big[ {\cal I}_{{\rm SAR},{\rm W}}(\by^S) \big] =
{\rm C}_{\rm W},\quad \quad 
{\rm C}_{\rm W} = \frac{\sigma_{\rm W}^2  a B}{2^{17/2} \pi^3 k_o^2 L^2},
\end{equation}
and the covariance is
\begin{align}
{\rm Cov} \big( {\cal I}_{{\rm SAR},{\rm W}}(\by^S),{\cal I}_{{\rm SAR},{\rm W}}({\by^S}') \big)
&={\rm C}_{\rm W}^2
\exp\left[ -\frac{2 (\byl^S-{\byl^S}')^2}{(c/B)^2} - \frac{(\byp^S-{\byp^S}')^2}{[L/(k_o a)]^2}\right],
\label{eq:covarSARW}
\end{align}
where we have used the fourth moment property satisfied by the Gaussian process
\begin{align}
\nonumber
&\EE \big[ \overline{\hat{W}}(\omega,\xp) \hat{W}(\omega',\xp') {\hat{W}}(\tilde{\omega},\tilde{\xp})  \overline{\hat{W}}(\tilde{\omega}',\tilde{\xp}') \big]\\
\nonumber
& = \sigma_{\rm W}^4 \big[ \delta(\omega - \omega')
\delta( \xp - \xp')  \delta(\tilde{\omega} - \tilde{\omega}')
\delta(\tilde{\xp} - \tilde{\xp}' )  \\
&\quad +\delta(\omega - \tilde{\omega})
\delta(\xp-\tilde{\xp} ) \delta( \omega' - \tilde{\omega}')
\delta(\xp'-\tilde{\xp}') 
\big] 
.
\label{eq:momfour}
\end{align}
By letting $\by^S = {\by^S}'$ in the last equation we obtain the variance
\begin{equation}
{\rm Var} \big( {\cal I}_{{\rm SAR},{\rm W}}(\by^S) \big) =
{\rm C}_{\rm W}^2 .
\end{equation}

The decay of the covariance \eqref{eq:covarSARW} shows that the noise induced speckle pattern has correlation radius $\lambda_oL/a$  in the cross range direction and $c/B$ in the range direction, which means that the image displays  bright and dark spots with these typical sizes. This is in addition to the random fluctuations induced by scattering in the random 
medium.

\subsection{Analysis of the two-point CINT imaging function}
\label{sect:analCINT}
We now describe the mean and variance of the two-point CINT imaging function 
\begin{align}
\nonumber
{\cal I}(\by^S,{\by^S}') = &
\frac{1}{(2\pi)^2} \int_{\RR} d \om  \int_{\RR}
d \om'  \int_{\RR} d \xp \int_{\RR} d \xp' \, \overline{\hat{R}(\omega,\xp)} {\hat{R}}(\omega',\xp') {\hat F}(\omega,\xp,\by^S) \nonumber \\
&\hspace{-0.2in}\times \overline{\hat F(\omega',\xp',{\by^S}')} \exp\Big[-\frac{(\xp-\xp')^2}{2X^2}- \frac{(\omega-\omega')^2}{2\Omega^2} - \frac{\xp^2+(\xp')^2}{a^2} \Big],
\label{def:cintgCont}
\end{align}
 which is related to CINT by \eqref{eq:HCINT_CINT}.   Its use 
in the high-resolution CINT imaging method is discussed in section \ref{sect:phaseRet}.

\subsubsection{The mean}
\label{sect:meanCINT}
The expression of the mean  is obtained from definitions \eqref{def:cintg}, \eqref{eq:greenhomo},
\eqref{eq:filterH} and equations (\ref{eq:paraxAp}--\ref{eq:hatD}). The calculation is the same in the homogeneous and the 
random medium, except that in the latter case we use the  moment formula \eqref{eq:secondmomentgreen} and in the 
former case there is no need for the expectation if there is no noise. 

We obtain that  in the noiseless case 
\begin{align}
\nonumber
\EE \big[ {\cal I}(\by^S, {\by^S}') \big]
={\rm C}  &\int_{\RR^2} d \by \int_{\RR^2} d \by' \, \rho(\by) \rho(\by')  {\cal K}^{(1)}_{\tilde{X},\tilde{\Omega} } \Big( \frac{\by^S+{\by^S}'}{2}-\frac{\by+\by'}{2}\Big)\\
&\times
{\cal K}^{(2)}_{a,B}\big( (\by^S-{\by^S}')-(\by-\by') \big) , \label{eq:mean2CINT}
\end{align}
with the same constant ${\rm C}$ as in \eqref{eq:SARhom} and with kernels
\begin{align}
{\cal K}^{(1)}_{\tilde{X},\tilde{\Omega}} (\by) =&  \pi \tilde{X}\tilde{\Omega}  \exp \left[ - \frac{2 \byp^2}{[L/(k_o \tilde X)]^2} -\frac{2 \byl^2}{(c/\tilde \Omega)^2}   \right]  ,\label{eq:kern1}\\
{\cal K}^{(2)}_{a,B}(\by) =&  \pi a B  \exp \left[ - \frac{\byp^2}{2 [L/(k_o a)]^2}
-\frac{\byl^2}{2 (c/B)^2}   
-2i k_o \byl \right]. \label{eq:kern2}
\end{align}
The first kernel gives the resolution  in the central spatial variable, modeled by the 
decay of \eqref{eq:kern1} on the scale $\la_o L/\tilde X$ in the cross-range direction and 
$c/\tilde \Omega$ in the range direction. If the medium were homogeneous,  
$\tilde{X}$ and $\tilde{\Omega}$ would depend on the window parameters $X$ and $\Omega$
in the definition \eqref{def:cintg} of ${\cal I}(\by^S,{\by^S}')$,  the aperture $a$ and the bandwidth $B$ 
of the probing pulse as follows
\begin{equation}
\frac{1}{\tilde{\Omega}^2}=\frac{1}{\Omega^2}+\frac{1}{B^2},\quad\quad 
\frac{1}{\tilde{X}^2}=\frac{1}{X^2}+\frac{1}{a^2}. 
\end{equation}
In the random medium, they also depend on the decoherence length $X_d$ and frequency $\Omega_d$ defined in  \eqref{eq:defXc},
\begin{equation}
\frac{1}{\tilde{\Omega}^2}=\frac{1}{\Omega_d^2} + \frac{1}{\Omega^2}+\frac{1}{B^2},\quad\quad 
\frac{1}{\tilde{X}^2}=\frac{1}{X_d^2} + \frac{1}{X^2}+\frac{1}{a^2}.
\end{equation}
The second kernel \eqref{eq:kern2} gives the resolution in the spatial offset, modeled by the 
decay on the scale $\la_o L/a$ in the cross-range direction and $c/B$ in the range direction.
These scales coincide with the resolution limits of the SAR imaging function in the homogeneous medium.

We conclude that the mean of the two-point CINT image displays excellent resolution  in 
the spatial offset locations $\by^S-{\by^S}'$, and reduced  resolution in the mid-point locations $(\by^S+{\by^S}')/2$.  
The latter is the same as the resolution of the CINT imaging function, obtained from \eqref{eq:HCINT_CINT} by setting $\by^S = {\by^S}'$ in  equation  \eqref{eq:mean2CINT},
\begin{align}
\EE \big[ {\cal I}_{\rm CINT}(\by^S) \big]
={\rm C}  &\int_{\RR^2} d \by \int_{\RR^2} d \by' \, \rho(\by) \rho(\by')  {\cal K}^{(1)}_{\tilde{X},\tilde{\Omega} } \Big( \by^S-\frac{\by+\by'}{2}\Big) {\cal K}^{(2)}_{a,B}(\by-\by').  \label{eq:meanCINT}
\end{align}
In the particular case of a single point scatterer, the CINT imaging function has the simple expression
\begin{equation}
 \EE \big[ {\cal I}_{\rm CINT}(\by^S) \big]
={\rm C}  \pi a B \rho_\star ^2 {\cal K}^{(1)}_{\tilde{X},\tilde{\Omega} } ( \by^S-\by_\star), \quad \rho(\by) = \rho_\star 
\delta(\by-\by_\star),  \label{eq:meanCINT1point}
\end{equation}
and its resolution can be compared easily to that of the mean SAR image in \eqref{eq:mean point}. We note in particular 
that the smaller the window parameter $\Omega$ and $X$ are, the worse the resolution of CINT is.  
We show in the next section that to ensure the statistical stability of the two-point CINT (and therefore of the CINT) image, the window parameters should satisfy 
\begin{equation}
X \lesssim X_d, \quad \Omega \lesssim \Omega_d.
\label{eq:goodXOm}
\end{equation} 

If scattering in the random medium is so weak that the wave components remain correlated across the aperture and bandwidth,
in the sense that $X_d > a$ and $\Omega_d > B$,
 then we can remove the windowing in \eqref{def:cintg} to obtain 
\[
\EE\big[ {\cal I}(\by^S,{\by^S}') \big]^2 \simeq \EE\big[ {\cal I}_{\rm SAR} (\by^S) \big]\EE\big[ {\cal I}_{\rm SAR} ({\by^S}') \big] ,
\]
and $ \EE\big[ {\cal I}_{\rm SAR} (\by^S) \big] $ is approximately given by its expression (\ref{eq:SARhom}) in homogeneous medium.

We are interested in strong scattering in the random medium, where 
$X_d \ll a$ and $\Omega_d \ll B$, and where the windowing in \eqref{def:cintg} is needed. The CINT image is then much blurrier than what the SAR image gives in homogeneus medium. 
This is the cost of statistical stability, as explained next. 

\subsubsection{The variance}
\label{sect:varCINT}
The calculation of the variance of the two-point CINT imaging function uses the fourth order moment formula 
\eqref{eq:momfour1}  for points and  frequencies satisfying
\begin{eqnarray*}
&&\omega_1 = \omega + \frac{\tilde{\omega}}{2}, \quad \omega_2= \omega -
\frac{\tilde{\omega}}{2}, \quad \quad \omega_3 = \omega' + \frac{\tilde{\omega}'}{2},
\quad \omega_4= \omega' - \frac{\tilde{\omega}'}{2},\\
 &&\bx_1 = \bx + \frac{\tilde{\bx}}{2}, \quad \bx_2= \bx -
\frac{\tilde{\bx}}{2}, \quad \quad \bx_3 = \bx' + \frac{\tilde{\bx}'}{2}, \quad
\bx_4= \bx' - \frac{\tilde{\bx}'}{2}.
\end{eqnarray*}
The center frequencies are  $\omega,\omega' \sim \omega_o$ and the frequency offsets satisfy $|\tilde{\omega}|, |\tilde{\omega}'|\lesssim \Omega$. Similarly, $\bx,
\bx' $ are center points in the aperture i.e., with range coordinate $L$ and cross-range coordinates of the order of $a$ and the   spatial offsets  $\tilde{\bx} = (0,\tilde x_\perp)$ and $ \tilde{\bx}' = (0,\tilde x_\perp')$ satisfy   $|\tilde x_\perp|, 
|\tilde x_\perp'| \lesssim 2 X$.  We are interested in the choice \eqref{eq:goodXOm} of the window parameters, where the two-point CINT imaging function is statistically stable, as shown below.  Definitions \eqref{eq:defXc}, \eqref{eq:defTau} and the assumption \eqref{eq:largeTau} give that $X < X_d \ll \ell_c$, and after long but straightforward calculations we obtain that in this regime the variance takes the simple form
\begin{align}
{\rm Var}\big( {\cal I} (\by^S, {\by^S}') \big) = O\Big( \frac{X^2}{X_d^2} +\frac{\Omega^2}{\Omega_d^2}\Big) \EE\big[ {\cal I} (\by^S, {\by^S}') \big]^2  .
\label{eq:varCINT}
\end{align}

We conclude that the two-point CINT imaging function gives statistically stable results when the window parameters satisfy the relation \eqref{eq:goodXOm} because then the coefficient of variation is smaller than one:
\begin{equation}
\mathcal{V}(\by^S,{\by^S}') = \frac{\sqrt{{\rm Var}\big( {\cal I}(\by^S,{\by^S}') \big)}}{\EE\big[ {\cal I}(\by^S,{\by^S}') \big] }< 1.
\label{eq:variationCINT}
\end{equation}
The optimal choice of the window parameters reflects the trade-off between 
the stability and resolution and corresponds to $X \lesssim X_d$ and $\Omega \lesssim \Omega_d$, as stated in \cite{borcea11,borcea06}.

\vspace{0.05in}
\begin{remark}
\label{rem.1}
We assumed a search  (imaging) region ${\cal D}$ of radius less than $\ell_c$ in order to simplify the expressions of the second- and fourth-order moments of the Green's function, and therefore  the mean and variance of the two-point CINT imaging function.  We have seen that the reflectivity function can be localized and imaged by CINT  at the scales $\lambda _o L / X$ and $c_o/\Omega$ in the cross-range and range directions.
This gives a consistent and relevant result because the second condition in (\ref{eq:condrtt})  and definition 
\eqref{eq:defXc} ensure that 
with $X \sim X_d$ and $\Omega \sim \Omega_d$ we have $\lambda _o L / X\ll \ell_c$ and $c/\Omega \ll \ell_c$ .
\end{remark}

\subsubsection{Additive noise}
\label{sect:CINTNoise}
The effect of the additive noise on the two-point CINT imaging function is modeled by  the expression
\begin{align}
\nonumber
{\cal I}_{\rm W}(\by^S,{\by^S}') = &
\frac{1}{(2\pi)^2} \int_{\RR} d \om  \int_{\RR}
d \om'  \int_{\RR} d \xp \int_{\RR} d \xp' \, \overline{\hat{W}(\omega,\xp)} {\hat{W}}(\omega',\xp') {\hat F}(\omega,\xp,\by^S) \nonumber \\
&\hspace{-0.25in}\times \overline{\hat F(\omega',\xp',{\by^S}')} \exp\Big[-\frac{(\xp-\xp')^2}{2X^2}- \frac{(\omega-\omega')^2}{2\Omega^2} - \frac{\xp^2+(\xp')^2}{a^2} \Big].
\label{def:cintWCont}
\end{align}
Using definitions \eqref{eq:covarW}, \eqref{eq:greenhomo}, \eqref{eq:filterH} 
and the paraxial approximation \eqref{eq:paraxAp}, we obtain the  mean
\begin{align}
\EE\big[ {\cal I}_{\rm W}(\by^S,{\by^S}') \big]  =
{\rm C}_{\rm W} 
\exp\left[ - \frac{ (\byl^S-{\byl^S}')^2}{(c/B)^2} - \frac{(\byp^S-{\byp^S}')^2}{2 [L/(k_o a)]^2}- 2i k_o(\byl^S-{\byl^S}') \right],
\label{eq:meanCINT_W}
\end{align}
with the same constant ${\rm C}_{\rm W}$ as in \eqref{def:Cn}.  The covariance is calculated using the Gaussian property (\ref{eq:momfour}) of the noise and the result is 
\begin{align}
{\rm Cov}\big(  {\cal I}_{\rm W}(\by^S,{\by^S}') , &\, {\cal I}_{\rm W}(\bz^S,  {\bz^S}') \big) 
={\rm C}_{\rm W}^2 \nonumber \\
& \times 
\exp\Big[ -\frac{(\byl^S- \bzl^S)^2}{(c/B)^2}- \frac{(\byp^S- \bzp^S)^2}{2 [L/(k_o a)]^2}
+ 2i k_o(\byl^S-\bzl^S)\Big] \nonumber \\
& \times \exp\Big[  -\frac{ ({\byl^S{}}' -{\bzl^S{}}')^2}{(c/B)^2}- \frac{({\byp^S{}}'-{\bzp^S}')^2}{2 [L/(k_o a)]^2}
- 2i k_o ({\byl^S}'-{\bzl^S}')\Big] .
\label{eq:covCINT_W}
\end{align}

When we set $\by^S = {\by^S}'$ and $\bz^S = {\bz^S}'$ in equations (\ref{eq:meanCINT_W}--\ref{eq:covCINT_W})
we obtain that  the additive noise effect on the classic CINT image
consists of speckle with uniform mean
\begin{align}
\EE\big[ {\cal I}_{{\rm CINT},{\rm W}}(\by^S) \big]  ={\rm C}_{\rm W},
\end{align}
and with covariance
\begin{align}
{\rm Cov}\big(  {\cal I}_{{\rm CINT},{\rm W}}(\by^S) , \, &{\cal I}_{{\rm CINT},{\rm W}}(\bz^S)\big) ={\rm C}_{\rm W}^2 \nonumber \\
&\times 
\exp\Big[ -\frac{2(\byl^S- \bzl^S)^2}{(c/B)^2}- \frac{(\byp^S- \bzp^S)^2}{[L/(k_o a)]^2}
+ 2i k_o(\byl^S-\bzl^S)\Big].
\label{eq:covCINTC_W}
\end{align}
Therefore, the noise induced speckle size in the CINT image is of the order of $\la_o L/a$ in the cross-range direction and $c/B$ in the range direction. These are smaller  than the cross-range resolution
$\la_o L/\tilde{X}$  and range resolution $c/\tilde \Omega$ obtained in the previous section. Thus, if the noise is weak (i.e., ${\rm C}_{\rm W}$ is small), then it does not affect the CINT image. If it is moderate, then it is possible to remove the induced 
speckle, up to the uniform mean,  using a low-pass filter on the CINT image.

However, the noise induced speckle plays a role in the two-point CINT image, 
because its typical size is of the same order as the  resolution in the offset spatial variables. We discuss this point further in the next section.

\section{High-resolution CINT imaging}
\label{sect:phaseRet}
With the window parameters chosen optimally, as explained above, so that the two-point CINT image is approximated 
by its mean, we obtain from equation \eqref{eq:mean2CINT} and definitions (\ref{def:HCINT}--\ref{eq:HatHCINT})  that  the high-resolution CINT image in the noiseless case is 
\begin{align}
\nonumber
\widehat{\cal I}_{\rm HCINT}(\bk{-2\bk_o}) & \approx \EE \big[ \widehat{\cal I}_{\rm HCINT}(\bk{-2\bk_o}) \big] \\
&=
{\rm C} \pi^4\Big(\frac{Lc}{k_o}\Big)^2|\hat{\rho}(\bk  -2\bk_o)|^2 \exp
\left[ -\frac{\bkp^2}{2 (a k_o/L)^2 } 
-\frac{\bkl^2}{2 (B/c)^2} \right],
\label{eq:goodPeak}
\end{align}
where $\bk = (\bkl,\bkp)$, $\bk_o = (k_o,0)$ and $\hat \rho$ is the Fourier transform of the unknown reflectivity function.
 Note that
 \begin{equation}
 \hat{\rho}(\bk  -2\bk_o) =\hat{\rho}_{k_o}(\bk),
 \label{eq:rhohat1}
 \end{equation}
 the Fourier transform of the reflectivity $\rho$ modulated in range  at wavenumber $k_o$, 
\begin{equation}
 \rho_{k_o}(\by)= \rho(\by) \exp(  2 i k_o \byl).
\label{eq:rhohat2}
\end{equation}

Equation \eqref{eq:goodPeak} shows that we can estimate $|\hat{\rho}_{k_o}(\bk)|$ at wave vectors 
$\bk = (\bkl,\bkp)$ with $|\bkl| \lesssim B/c$ and $|\bkp| \lesssim a/(\la_o L)$.  In the spatial domain, this corresponds to 
sampling $\rho$ on a grid of size $\la_o L/a$ in the cross-range direction and $c/B$ in the range direction.  
An estimate of $\rho$ on such a grid can be obtained  from \eqref{eq:goodPeak} using phase retrieval, as explained in section \ref{sect:HRPRA}

In principle,  there may be another way of  estimating $\rho$ from the two-point CINT image, without phase retrieval. 
In definition (\ref{def:HCINT}--\ref{eq:HatHCINT}) of HCINT we integrate  ${\cal I}(\by^S,{\by^S}')$ over the center points
$(\by^S + {\by^S}')/2$ and then take the Fourier transform with respect to the offset $\by^S-{\by^S}' $. We could consider 
instead the function
\begin{align}
\hat {\mathfrak I}(\bk-2 \bk_0,\tilde \bk) = \int_{\RR^2} d {\by}^S \int_{\RR^2} d \tilde{\by}^S \, {\cal I}\Big( {\by}^S+ \frac{\tilde{\by}^S}{2},
 {\by}^S- \frac{\tilde{\by}^S}{2}\Big) e^{-i \tilde \bk \cdot {\by}^S - i \bk \cdot \tilde \by^S} ,
\end{align}
and obtain from \eqref{eq:mean2CINT} and the definitions (\ref{eq:kern1}--\ref{eq:kern2}) that 
\begin{align}
 \hat {\mathfrak{I}}(\bk-2 \bk_0,\tilde \bk) \approx
\EE\big[ \hat {\mathfrak{I}}(\bk-2 \bk_0,\tilde \bk) \big] =
\hat \rho \Big(\bk-2 \bk_o + \frac{\tilde \bk}{2} \Big) \overline{\hat \rho \Big(\bk-2 \bk_o - \frac{\tilde \bk}{2} \Big) } \nonumber 
\\ \times {\rm C} \pi^4 \Big(\frac{Lc}{k_o}\Big)^2 \exp \left[ - \frac{\bkp^2}{2 (a k_o/L)^2} - \frac{\bkl^2}{2 (B/c)^2} - \frac{\tilde{\bkp}^2}{8 (\tilde{X} k_o/L)^2} - \frac{\tilde \bkl^2}{8 (\tilde \Omega/c)^2} \right].
\label{eq:altern1}
\end{align}
This can be used to determine the phase of $\hat \rho$. For instance, from \eqref{eq:altern1} we get 
$$
{\rm arg}\, \widehat{\mathfrak{I}}(\bk - 2\bk_o,\tilde \bk)   \approx 
{\rm arg} \,\hat{\rho}\big(\bk - 2\bk_o +\frac{\tilde \bk}{2}\big) - {\rm arg} \,\hat{\rho} \big(\bk - 2\bk_o - \frac{\tilde \bk}{2} \big),
$$
so we could estimate $\nabla_{\bk} {\rm arg} \hat{\rho}(\bk - 2\bk_o )$ from $\nabla_{\tilde{\bk}} {\rm arg} \, \hat {\mathfrak{I}} (\bk - 2\bk_o,\tilde \bk)\big|_{\tilde \bk = {\bf 0}}$ and then integrate to get the phase.
Our experience  is that this approach is not stable, because it requires 
an estimate of $\arg \hat \rho$  and its gradient over the whole domain of $\bk$, while  $\hat{\rho}$ 
may be  very small in some regions of this domain. 

The numerical results in section \ref{sect:numerics} are based on the 
phase retrieval method described next.

\subsection{Phase retrieval}
\label{sect:HRPRA}
The goal of phase retrieval   \cite{fienup78,fienup82,fienup87,fienup90,review15} is to determine a function 
$\eta(\bx)$ with Fourier transform $\hat \eta(\bk)$, such that 
\begin{align}
&|\hat \eta(\bk)| = |\hat \rho_{k_o}(\bk)|, \qquad \bk = (\bkl,\bkp), ~~ |\bkl| \lesssim \frac{B}{c}, ~~|\bkp| \lesssim \frac{a}{\la_o L},
\label{eq:etaPRA}
\end{align}
with $|\hat \rho_{k_o}(\bk)|$ obtained from (\ref{eq:goodPeak}--\ref{eq:rhohat1}). Note that the restriction of the wave vector 
$\bk$ to the domain in \eqref{eq:etaPRA} ensures that the exponential in the right-hand side of 
\eqref{eq:goodPeak} is of order one and thus can be safely inverted.
If  the reflectivity function is known to be nonnegative valued, then we 
can use a phase retrieval algorithm with positivity constraints 
\begin{equation}
\eta(\bx)\exp(-2 i k_o x_\parallel)  \ge 0, \label{eq:constraints}
\end{equation}
which is known to give a good reconstruction with resolution $\la_o L/a$ in the cross-range direction and $c/B$ in the range direction, up to a global shift and a symmetry  with respect to the origin \cite{fienup78,fienup82}. 
The global shift uncertainty can be compensated by the standard CINT image, with a precision given by the classic CINT resolution. That is to say, the estimated reflectivity 
\begin{equation}
\rho_{\rm est}(\bx) = \eta(\bx)\exp(-2 i k_o x_\parallel),
\end{equation}
can be centered in the zoom area within a peak of CINT, where we seek to improve the resolution of the image.

\vspace{0.05in}
\begin{remark}
\label{rem:HCINT1}
One could also apply a phase retrieval algorithm with support constraint \cite{fienup87} determined from the standard CINT image. In fact, it should be possible to apply the two-step approach suggested in  \cite{fienup90} to extract a complex valued $\rho$ from $|\hat \rho_{k_o}|$ and  the low resolution standard CINT image. Moreover, if  the support of $\rho$ is small, i.e. a few points, then one can use a phase retrieval algorithm with sparsity contraints \cite{review15}. The latter is more sensitive to noise than the phase retrieval  with positivity constraints used in our numerical simulations.
\end{remark}

\subsection{Additive noise}
\label{sect:AddNoiseHCINT}
The additive noise contribution to the HCINT imaging function is, in the mean, 
\begin{equation}
\EE\big[ \hat{\cal I}_{\rm HCINT,W}(\bk{-2\bk_o} ) \big]  =\frac{\sqrt{2}\pi {\rm C}_{\rm W} |\cal D|  c L}{a k_o B}
\exp\Big[ - \frac{\bkp^2}{ 2 (a k_o/L)^2} -\frac{ \bkl^2 }{4 (B/c)^2} \Big],
\label{eq:peakhatHCINTadd}
\end{equation}
where $|\cal D|$ is the area of the imaging region.  Note that \eqref{eq:peakhatHCINTadd} peaks  at $\bk = 0$, just as \eqref{eq:goodPeak}, and has similar decay. This makes it difficult to filter out the noise effect, which will impede the high-resolution imaging via phase retrieval when the noise is strong enough. 

%------------------------------------

\section{Numerical results}
\label{sect:numerics}
\begin{figure}
\centerline{
 \includegraphics[width=6.cm]{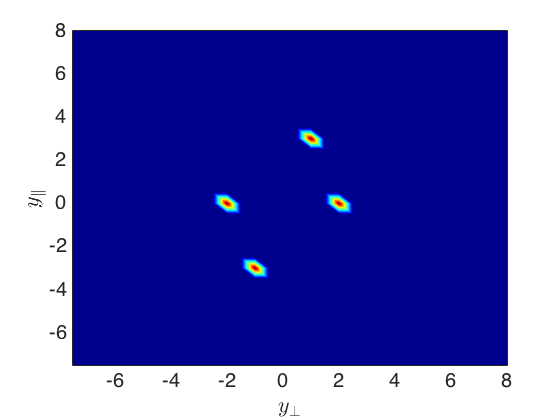} 
 }
\caption{The reflectivity model of four identical point-like  scatterers. The abscissa is cross-range and 
the ordinate is range, in multiples of the central wavelength $\lambda_o$. }
\label{fig:simu0}
\end{figure}

In this section we use numerical simulations to illustrate the performance of the HCINT imaging method.  
We consider a  reflectivity $\rho$ supported at four identical point-like scatterers, as in Fig. \ref{fig:simu0}.
The data \eqref{eq:data} are  generated with the random travel time model\footnote{In the analysis we assumed that 
$\ell_c \ll L$ in order to use the central limit theorem and  obtain Gaussian random travel time fluctuations ${\cal T}_\mu$. In the numerical simulations we use a Gaussian $\mu$, which means that ${\cal T}_\mu$ is Gaussian, no matter 
the ratio $L/\ell_c$. } described in section \ref{sect:model}, using a Gaussian zero-mean random process $\mu$ for the fluctuations of the wave speed, with correlation length   $\ell_c=L$ and  $\sigma=0.06$. The additive noise 
has standard deviation $\sigma_{\rm W}$ equal to  $0\%$, $20\%$ or $40\%$ of the maximal amplitude of the returns.
The bandwidth is $B/\om_o=1/5$ and $N=60$. The window parameters of the CINT function are $\Omega=B/5$ and $X=a/5$.
The imaging functions are calculated as defined in sections \ref{sect:imSAR}--\ref{sect:HCINT} and the
phase-retrieval is done with the simple (error-reduction) algorithm proposed in \cite{fienup87}.
There are better phase retrieval algorithms but this simple one was sufficient for our purpose.

\begin{figure}
\centerline{
\begin{tabular}{cc}
 \includegraphics[width=6.cm]{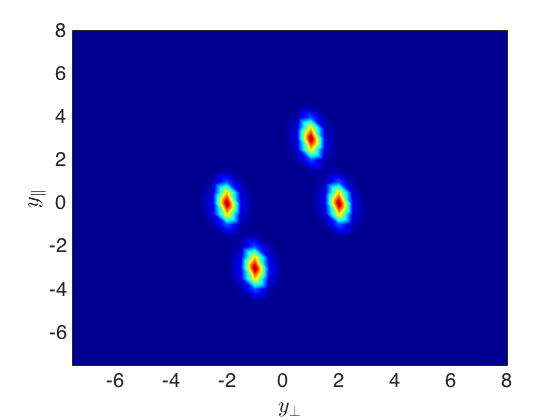} 
&
 \includegraphics[width=6.cm]{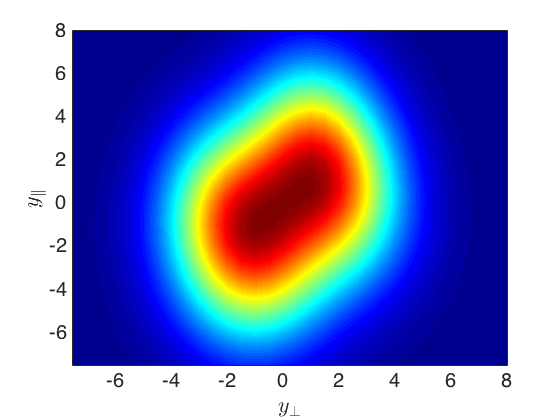} \\
{\bf (a)}: ${\cal I}_{\rm SAR}$ &
{\bf (b)}: ${\cal I}_{\rm CINT}$\\
 \includegraphics[width=6.cm]{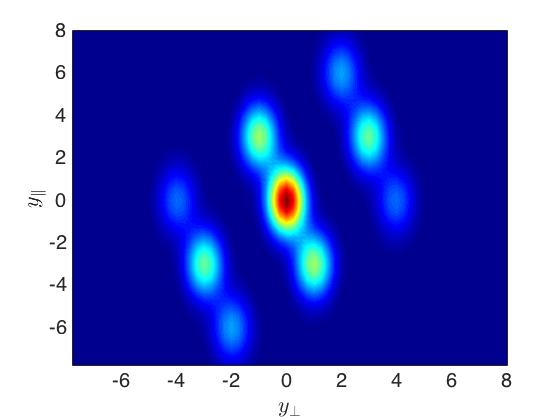} 
&
 \includegraphics[width=6.cm]{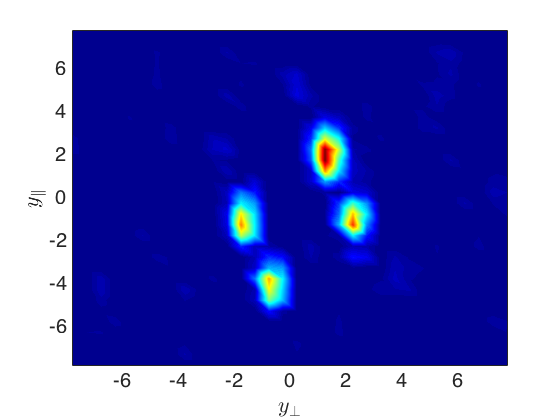} \\
{\bf (c)}: ${\cal I}_{\rm HCINT}$ &
{\bf (d)}: Estimated $\rho$\\
 \end{tabular}
 }
\caption{The imaging functions: ${\cal I}_{\rm SAR}$ (a),
${\cal I}_{\rm CINT}$ (b), ${\cal I}_{\rm HCINT}$ (c), and the reconstructed reflectivity (d).
The axes are as in Fig. \ref{fig:simu0}. No medium fluctuations and no additive noise.}
\label{fig:simu00}
\end{figure}

\begin{figure}
\centerline{
\begin{tabular}{cc}
 \includegraphics[width=6.cm]{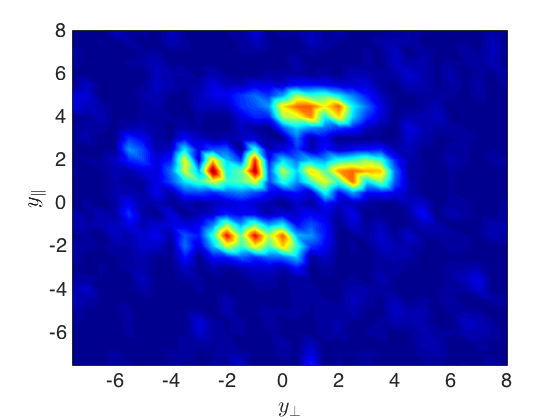} 
&
 \includegraphics[width=6.cm]{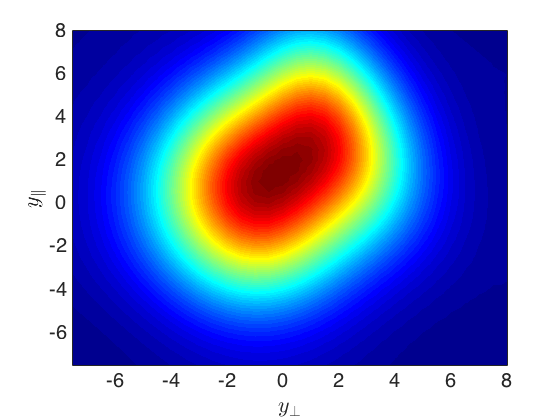} \\
{\bf (a)}: ${\cal I}_{\rm SAR}$ &
{\bf (b)}: ${\cal I}_{\rm CINT}$\\
 \includegraphics[width=6.cm]{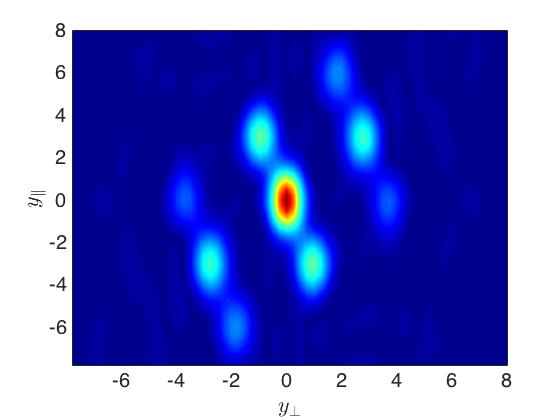} 
&
 \includegraphics[width=6.cm]{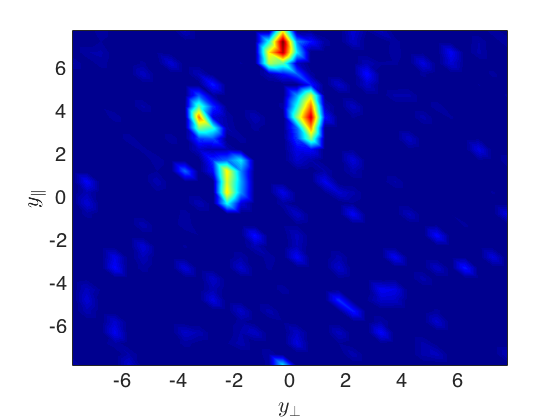} \\
{\bf (c)}: ${\cal I}_{\rm HCINT}$ &
{\bf (d)}: Estimated $\rho$
 \end{tabular}
 }
\caption{The imaging functions: ${\cal I}_{\rm SAR}$ (a),
${\cal I}_{\rm CINT}$ (b), ${\cal I}_{\rm HCINT}$ (c), and the reconstructed reflectivity (d).
The axes are as in Fig. \ref{fig:simu0}. 
Here there moderate additive noise ($20\%$) and strong medium perturbations, corresponding to $\om_o \tau = 6 \pi$ (recall \eqref{eq:largeTau}).}
\label{fig:simu248}
\end{figure}

\begin{figure}
\centerline{
\begin{tabular}{cc}
 \includegraphics[width=6.cm]{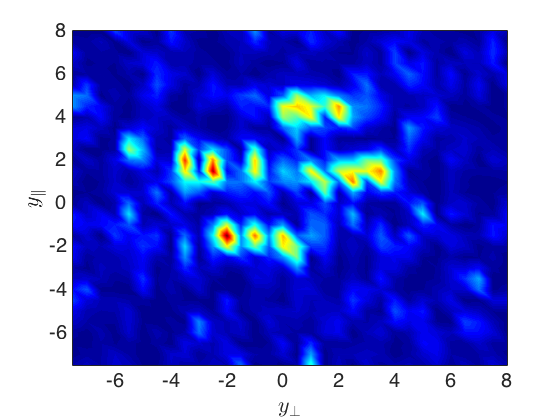} 
&
 \includegraphics[width=6.cm]{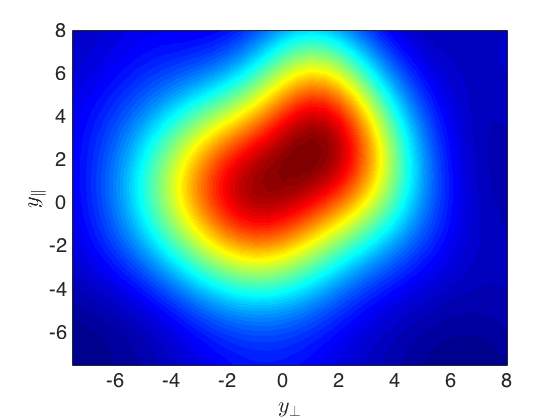} \\
{\bf (a)}: ${\cal I}_{\rm SAR}$ &
{\bf (b)}: ${\cal I}_{\rm CINT}$\\
 \includegraphics[width=6.cm]{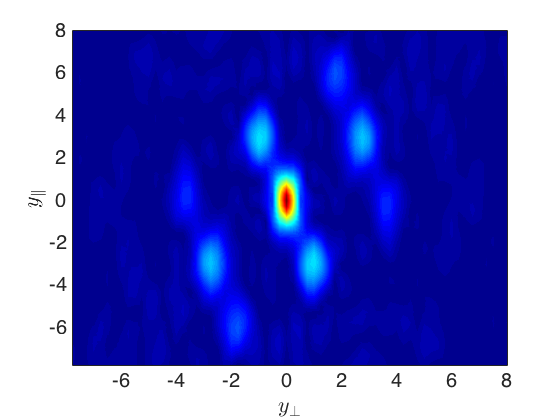} 
&
 \includegraphics[width=6.cm]{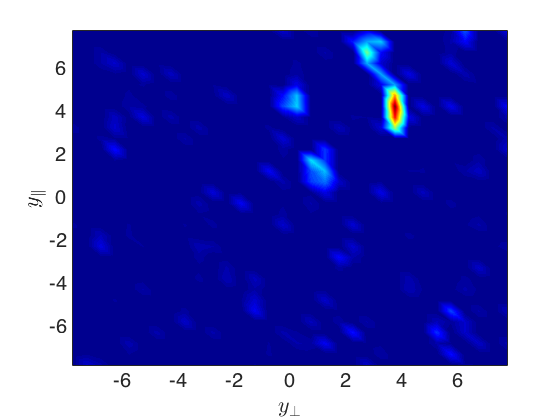} \\
{\bf (c)}: ${\cal I}_{\rm HCINT}$ &
{\bf (d)}: Estimated $\rho$\\
 \includegraphics[width=6.cm]{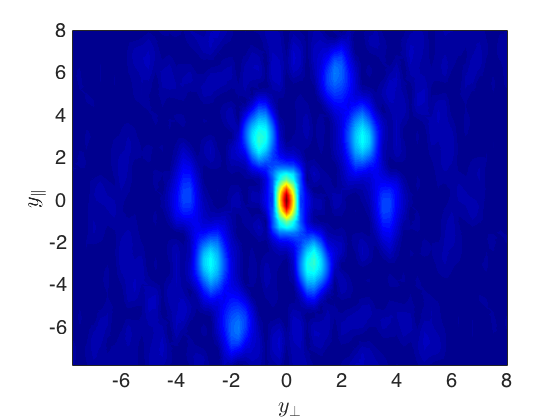} 
&
 \includegraphics[width=6.cm]{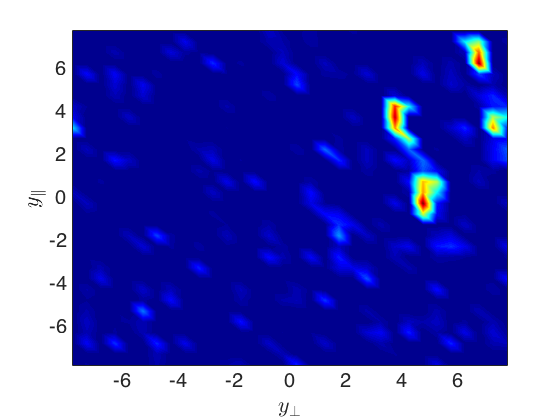} \\
{\bf (e)}: ${\cal I}_{\rm HCINT}$ modified&
{\bf (f)}: estimated $\rho$ modified
 \end{tabular}
 }
\caption{The imaging functions: ${\cal I}_{\rm SAR}$ (a),
${\cal I}_{\rm CINT}$ (b), ${\cal I}_{\rm HCINT}$ (c), and the reconstructed reflectivity (d).
The axes are as in Fig. \ref{fig:simu0}. 
Strong additive noise ($40\%$) and strong medium perturbations, corresponding to $\om_o \tau = 6 \pi$.
In picture (e) the main peak of the function ${\cal I}_{\rm HCINT}$ has been reduced by $20\%$. The resulting estimate of the 
reflectivity is plotted in picture (f).
}
\label{fig:simu448}
\end{figure}
The SAR imaging function gives a good estimate of the support of $\rho$ in the absence of noise  (Fig. \ref{fig:simu00}(a)) 
but it performs poorly in the presence of medium perturbations (Fig. \ref{fig:simu248}(a)--\ref{fig:simu448}(a)).
The standard CINT function gives a robust but low-resolution  image, in the absence or in the presence
of medium perturbations and additive noise (Fig. \ref{fig:simu0}(b)--\ref{fig:simu448}(b)).
The HCINT imaging function is displayed in Fig. \ref{fig:simu0}(c)--\ref{fig:simu448}(c) and gives a high-resolution image, in the absence or in the presence
of medium perturbations (Fig. \ref{fig:simu0}(d)--\ref{fig:simu448}(d)).

However, HCINT  is sensitive to strong additive noise. The central peak of ${\cal I}_{\rm HCINT}(\by^S)$ is enhanced as predicted in section \ref{sect:AddNoiseHCINT} and as seen in Fig. \ref{fig:simu448}(c), where the central peak dominates the others, compared to the other (c) labeled pictures. Consequently, the phase-retrieval algorithm cannot determine the correct amplitudes of the four peaks of the function $\rho$ (Fig. \ref{fig:simu448}(d)).
By reducing the amplitude of the main peak of  ${\cal I}_{\rm HCINT}(\by^S)$  by $20\%$ we get a better image
Fig. \ref{fig:simu448}(f). In practice,
there may be a way to automate this procedure, based on an estimate of the noise level. In any case, the phase-retrieval algorithm is known to be sensitive to noise. 

As we stated earlier, there is ambiguity in the estimated reflectivity in the sense that $\rho(\bx)$ and $\rho(\bx-\bx_\star)$
cannot be distinguished, for arbitrary $\bx_\star$. In the figures we display the results given by the phase retrieval algorithm, 
but the reconstructions could be shifted by hand to the center of  the peak of the CINT image.
Since $\rho(\bx)$ and $\rho(-\bx)$ have the same modulus of the Fourier transform, it is impossible to distinguish them 
with phase retrieval. This is not an issue in Fig.  \ref{fig:simu0}(d)--\ref{fig:simu448}(d) because the true reflectivity is symmetric with respect to the origin.
\section{Summary}
\label{sect:sum}
We introduced a novel interferometric imaging method for high-resolution synthetic aperture imaging  of the reflectivity 
of a remote region, when the waves propagate through scattering random media. The method builds on the coherent interferometric (CINT) approach which uses empirical cross-correlations calculated over carefully  chosen data sets  in order to mitigate the distortion of the wave caused by scattering. 
This mitigation comes at the expense of the resolution. The algorithm introduced in this paper 
is based on a modified version of the CINT method, where the imaging scene is sampled at pairs of points. It shows how 
to use this modified imaging function  to estimate the modulus of the Fourier transform of the unknown reflectivity function.
The image of the reflectivity is obtained from this estimate using a phase retrieval algorithm, and  the resolution 
is comparable to that of imaging through known and non-scattering media.  The imaging method inherits the robustness of CINT with respect to the uncertainty of the random medium. However, the phase retrieval part is sensitive to noise. 

In its current form, the imaging method is   computationally intensive. However, the principle should be of interest for 
synthetic aperture radar and other imaging modalities, and it should be possible to find more efficient  implementations that limit the numerical overburden and can be used in three-dimensional setups.

\section*{Acknowledgements}
This research is supported in part by the Air Force Office of Scientific Research under award number FA9550-18-1-0131.


\begin{thebibliography}{9}

\bibitem{borcea11}
L. Borcea, J. Garnier, G. C. Papanicolaou, and C. Tsogka, 
Enhanced statistical stability in coherent interferometric imaging, 
Inverse Problems {\bf 27}, 085004 (2011).

\bibitem{borcea06} 
L. Borcea,  G. C.  Papanicolaou, and C. Tsogka, 
Adaptive interferometric imaging in clutter and optimal illumination,
Inverse Problems {\bf 22}, 1405--1436 (2006).

\bibitem{borcea18}
L. Borcea and I. Kocyigit,
Passive array imaging in random media,
IEEE Trans. on Computational Imaging {\bf 4}, 459--469 (2018).

\bibitem{cheney}
M. Cheney,
A mathematical tutorial on synthetic aperture radar,
SIAM Rev. {\bf 43}, 301--312 (2001).

\bibitem{curlander}
J. C. Curlander and R. N. McDonough, 
Synthetic aperture radar,
Wiley, New York, 1991.

\bibitem{fienup78}
J. R. Fienup, 
Reconstruction of an object from the modulus of its Fourier transform,
Opt. Lett. {\bf 3}, 27--29 (1978).

\bibitem{fienup82}
J. R. Fienup, 
Phase retrieval algorithms: a comparison, 
Appl. Opt. {\bf 21}, 2758--2769 (1982).

\bibitem{fienup87}
J. R. Fienup,
Reconstruction of a complex-valued object from the modulus of its Fourier transform using a support constraint,
J. Opt. Soc. Am. A {\bf 4}, 118--123 (1987).

\bibitem{fienup90}
J. R. Fienup and A. M. Kowalczyk,
Phase retrieval for a complex-valued object by using a low-resolution image,
J. Opt. Soc. Am. A {\bf 7}, 450-458 (1990).

\bibitem{noisebook}
J. Garnier and G. Papanicolaou,
Passive imaging with ambient noise,
Cambridge University Press, Cambridge, 2016.

\bibitem{garniersolna08}
J. Garnier and K. S\o lna,
Coherent interferometric imaging for synthetic aperture radar in the presence of noise,
Inverse Problems {\bf 24}, 055001 (2008).

\bibitem{GarnierSolna}
J. Garnier and K. S\o lna,
Fourth-moment analysis for wave propagation in the white-noise paraxial regime,
Archive for Rational Mechanics and Analysis {\bf 220}, 37--81 (2016).

\bibitem{ishimaru}
A. Ishimaru, 
Wave propagation and scattering in random media, Vol. 2, 
Academic press, New York, 1978. 

\bibitem{rytov}
S. M. Rytov, Y. A. Kravtsov, and V. I. Tatarskii, 
Principles of statistical radiophysics. 4. Wave Propagation through random media, 
Springer Verlag, Berlin, 1989.

\bibitem{review15}
Y. Shechtman, Y. C. Eldar, O. Cohen, H. N. Chapman, J. Miao, and M. Segev,
Phase retrieval with application to optical imaging: A contemporary overview,
 IEEE Signal Processing Magazine {\bf 32} 87-109 (2015).
 
\bibitem{tatarski}
V. I. Tatarski, 
Wave propagation in a turbulent medium, 
Dover, New York, 1961.

\bibitem{vanRoss}
M. C. W. van Rossum  and Th. M. Nieuwenhuizen, 
Multiple scattering of classical waves: microscopy, mesoscopy, and diffusion,
Reviews of Modern Physics  {\bf 71}, 313--370 (1999).

\end{thebibliography}
\end{document}